\documentclass[russian,english,12pt]{article}
\usepackage[english,russian]{babel}
\usepackage{babel,amsmath,amssymb,epsfig,float}
\usepackage[T2A]{fontenc}
\usepackage[cp1251]{inputenc}
\usepackage{graphicx}

\begin{document}
\selectlanguage{russian}

\begin{center}
{\LARGE\bf
Описание ультрахолодных атомов в одномерной геометрии гармонической ловушки с реалистическим взаимодействием}

\vspace{5mm}

И.С. Ишмухамедов, Д.С. Валиолда, С.А. Жаугашева\\
\vspace{2mm}
Объединенный институт ядерных исследований, Дубна \\
Казахский национальный университет им. аль-Фараби, Алматы

\end{center}
\vspace{5mm}

\begin{abstract}
В работе выполнен расчет энергии основного состояния двух атомов в одномерной геометрии гармонической оптической ловушки. Получена зависимость энергии от одномерной длины рассеяния, отвечающей различным интенсивностям потенциала межатомного взаимодействия $V_{int}(x)=V_0\exp{ \left\{-2cx^2\right\}}$. Расчет произведен численным и аналитическим методами. В качестве аналитического метода выбран метод осцилляторного представления (ОП), успешно применяемый для расчета связанных состояний различных малочастичных систем.
Основными результатами данной работы являются:
1) численное исследование области применимости в этой задаче ранее использованного приближения потенциала нулевого радиуса;
2) исследование границы применимости метода ОП для потенциала $V(x)=V_{conf}(x)+V_{int}(x)=\frac{x^2}{2}+V_0\exp{ \left\{-2cx^2\right\}}$.
\end{abstract}

\selectlanguage{english}
\begin{abstract}
We compute the ground state energy of two atoms in a one-dimensional geometry of a harmonic optical trap. We obtain a dependence of the energy on a one-dimensional scattering length, which corresponds to various strengths of the interaction potential $V_{int}(x)=V_0\exp{ \left\{-2cx^2\right\}}$. The calculation is performed by numerical and analytical methods. For the analytical method we choose the oscillator representation method (OR), which has been successfully applied to computations of bound states of various few-body systems. The main results of this paper are: 1) numerical investigation of the validity range of the previously used pseudopotential method; 2) investigation of the validity range of the OR for the potential $V(x)=V_{conf}(x)+V_{int}(x)=\frac{x^2}{2}+V_0\exp{ \left\{-2cx^2\right\}}$.
\end{abstract}

\selectlanguage{russian}

\section*{Введение}

Исследования по ультрахолодным атомам представляют интерес в связи с уникальной возможностью моделировать и управлять такими физическими явлениями как сверхпроводимость, сверхтекучесть \cite{Feshbach}, химические реакции с образованием молекул, применимых для элементов квантового компьютера \cite{JILA, Feshbach}, а также кварк-глюонная плазма, возникшая в первые моменты Большого Взрыва \cite{Turlapov, Feshbach}.
В этих экспериментах атомы находятся в условиях ограниченной геометрии, возникающей в результате взаимодействия атомов с внешним оптическим потенциалом \cite{Pitaevskii}. Примерами служат квазиодномерная и квазидвумерная геометрии оптических ловушек \cite{Melezhik}. В первом случае (соответствующем сигарообразному виду ловушки) мы имеем практически свободное движение атомов вдоль одной из осей, а их поперечное движение ограничено и квантованно. Во втором случае (в случае ловушки в форме "блина")  квантованной является только одна из осей.

Оптическая ловушка представляет собой стоячую оптическую волну, которую в нулевом приближении можно описать гармоническим потенциалом. В различных экспериментах с ультрахолодными атомами были обнаружены резонансы которые противоречат теории основанной на гармоническом приближении для потенциала ловушки \cite{Drummond}. Авторы \cite{Drummond} утверждают, что подобное отклонение теории от эксперимента обусловлено ангармоническими поправками. Однако вычисления ангармонических поправок, выполненных в данной работе, в первом порядке теории возмущений нельзя признать удовлетворительными \cite{Efimov}.

Данная статья представляет собой первый шаг на пути описания ангармонизма вне рамок теории возмущений, где мы вычисляем энергии основного состояния атомов в одномерном случае, т.е. когда потенциал межатомного взаимодействия и потенциал ловушки являются одномерными, последний аппроксимируется параболической функцией. Подобный расчет для случая межатомного взаимодействия вида дельта-функции (псевдопотенциал нулевого радиуса) был проведен в \cite{Busch}. Наш случай отличается тем, что мы используем реалистический потенциал гаусса. Обнаружено заметное отклонение от \cite{Busch} (Рис. 2). Отметим также расчет спектра для реалистического трехмерного потенциала межатомного взаимодействия и двумерного осцилляторного потенциала ловушки в \cite{Olshanii}, где было получено отклонение от результата, полученного в псевдопотенциальном подходе.

Мы используем как численный, так и аналитический подход к решению поставленной задачи. Численный расчет включает в себя расчет зависимости обратной длины рассеяния от параметров потенциала взаимодействия. Расчет энергии производится с использованием конечно-разностной аппроксимации второго порядка и метода обратной итерации. Аналитический подход расчета энергии осуществляется методом осцилляторного представления (ОП). Метод ОП основан на идеях и методах скалярной квантовой теории поля и его эффективность продемонстрирована при расчете связанных состояний различных малочастичных систем \cite{Efimov}

Статья построена следующим образом: в первом разделе приведены результаты расчетов зависимости обратной длины рассеяния от параметров потенциала взаимодействия и зависимости энергии основного состояния от обратной длины рассеяния. Полученные результаты сравниваются с расчетом, выполненным в псевдопотенциальном подходе \cite{Busch}; во втором разделе изложен формализм метода ОП применительно к данной задаче. Исследована область применения метода как функция параметров потенциала взаимодействия; в заключении обсуждаются полученные результаты.

\section{Энергия основного состояния}
Целью работы является исследование зависимости энергии основного уровня $E$ и соответствующей волновой функции (ВФ) $\Psi$ двух бозонных атомов, плененных в одномерной гармонической ловушке, от интенсивности межатомного взаимодействия. Эта задача сводится к решению одномерного уравнения Шредингера (УШ):
\begin{eqnarray}
\left\{ -\frac{1}{2}\frac{d^2}{dx^2}+\frac{x^2}{2}+V_0\exp{ \left\{-2cx^2\right\}}
\right\}\Psi(x)=E\Psi(x),
\end{eqnarray}
с нулевыми граничными условиями $\Psi(|x|\rightarrow \infty) \rightarrow 0$ для ВФ относительного движения атомов в гармоническом потенциале $V_{conf}(x)=\dfrac{x^2}{2}$. Здесь и далее, если не оговорено противное, мы используем осцилляторную систему единиц $m=\hbar=\omega=1$. Межатомное взаимодействие моделировалось двухпараметрическим гауссовым потенциалом $V_{int}(x)=V_0\exp{ \left\{-2cx^2\right\}}$\footnote[1] {коэффициент 2 в экспоненте выбран из соображения удобства.}, в котором параметр $V_0$ задает глубину, а $c$ определяет ширину потенциала. Мы исследовали зависимость $E$ и $\Psi(x)$ от обратной длины рассеяния $a^{-1}(V_0)$ на одномерном гауссовом потенциале $V_0\exp{ \left\{-2cx^2\right\}}$ в отсутствие удерживающего потенциала ловушки $V_{conf}$. Вначале была рассчитана зависимость $a^{-1}(V_0)$ при численном решении УШ (1) для случая $V_{conf}=0$ для состояния непрерывного спектра $E>0$ с граничным условием для ВФ:
\begin{eqnarray}
\lim_{|x|\rightarrow\infty}\Psi(x) \sim \cos(k|x|+\delta),
\end{eqnarray}
где $k=\sqrt{2E}$, а $\delta$ - фаза рассеяния. При этом одномерная длина рассеяния определяется как:
\begin{eqnarray}
a=\lim_{k \rightarrow 0} \frac{\cot(\delta)}{k}
\end{eqnarray}
Рассчитанная зависимость $a^{-1}(V_0)$, при различных параметрах $c$ потенциала взаимодействия, представлена на Рис. 1. Аналогичный расчет трехмерной длины рассеяния на короткодействующем потенциале был проведен в \cite{Melezhik2}.

\begin{center}
\begin{tabular}{cc}

\includegraphics*[width=90mm]{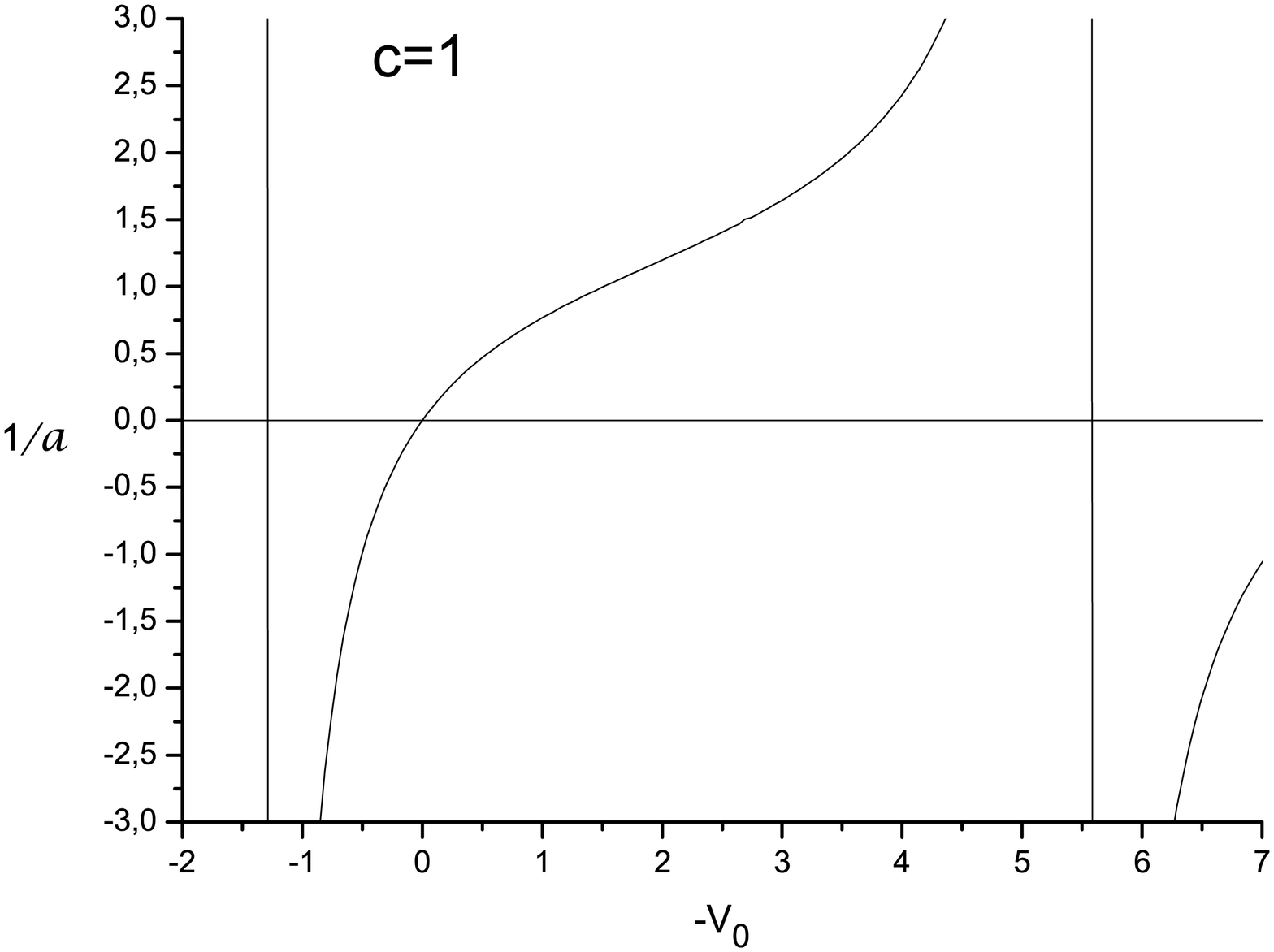} & \hspace{-16mm}
\includegraphics*[width=90mm]{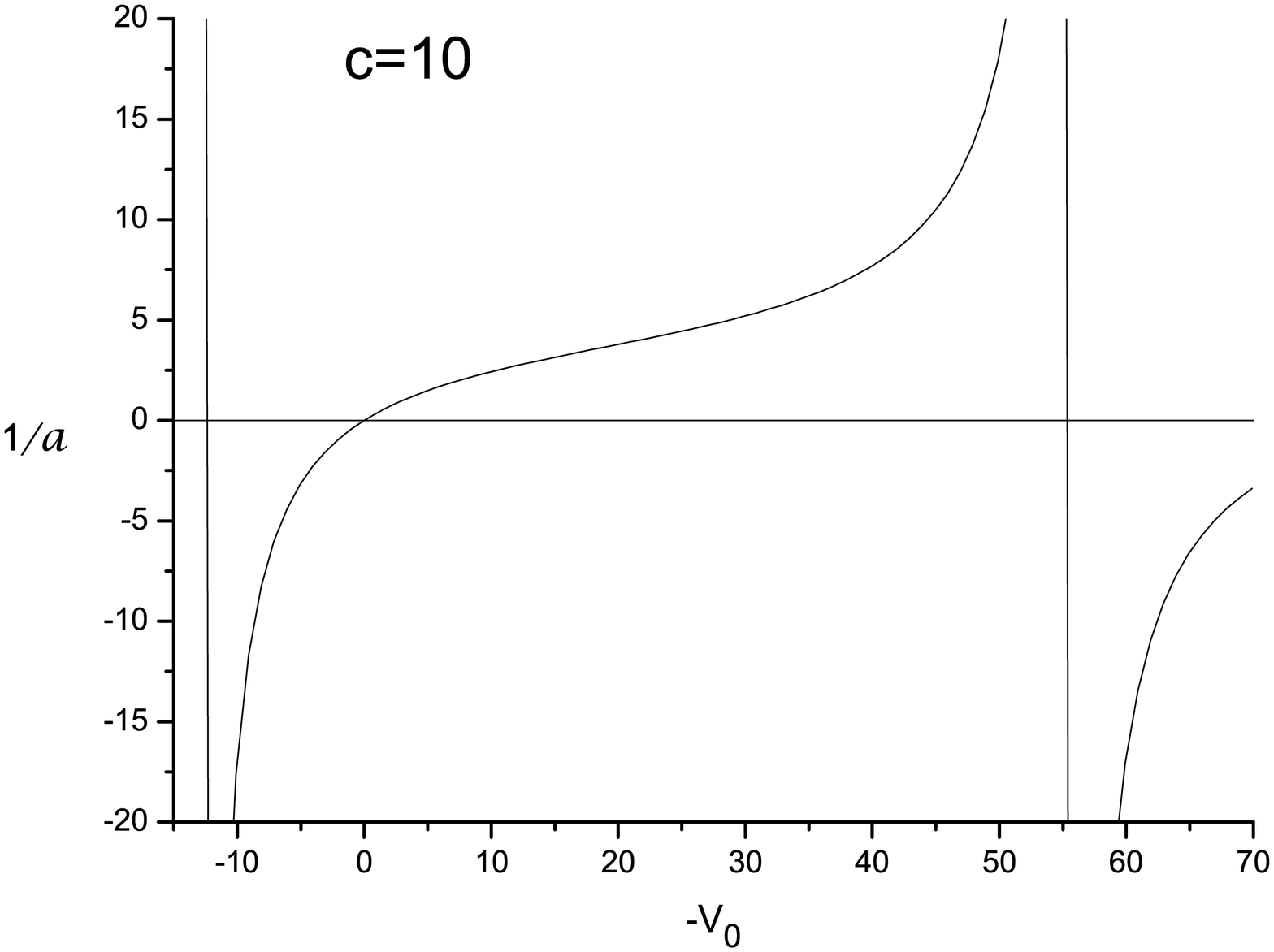} \\[-4mm]

\includegraphics*[width=90mm]{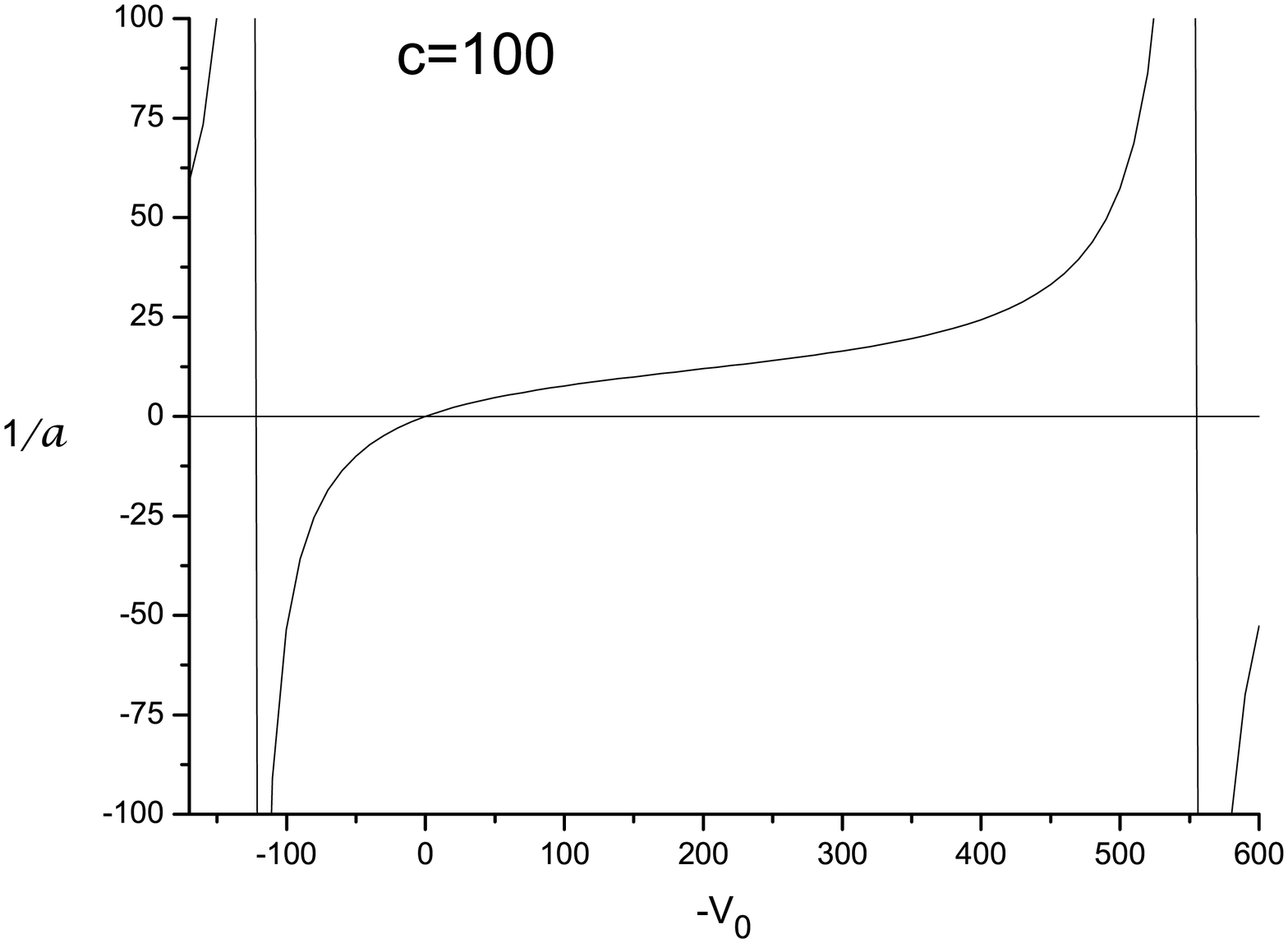} & \hspace{-16mm}
\includegraphics*[width=90mm]{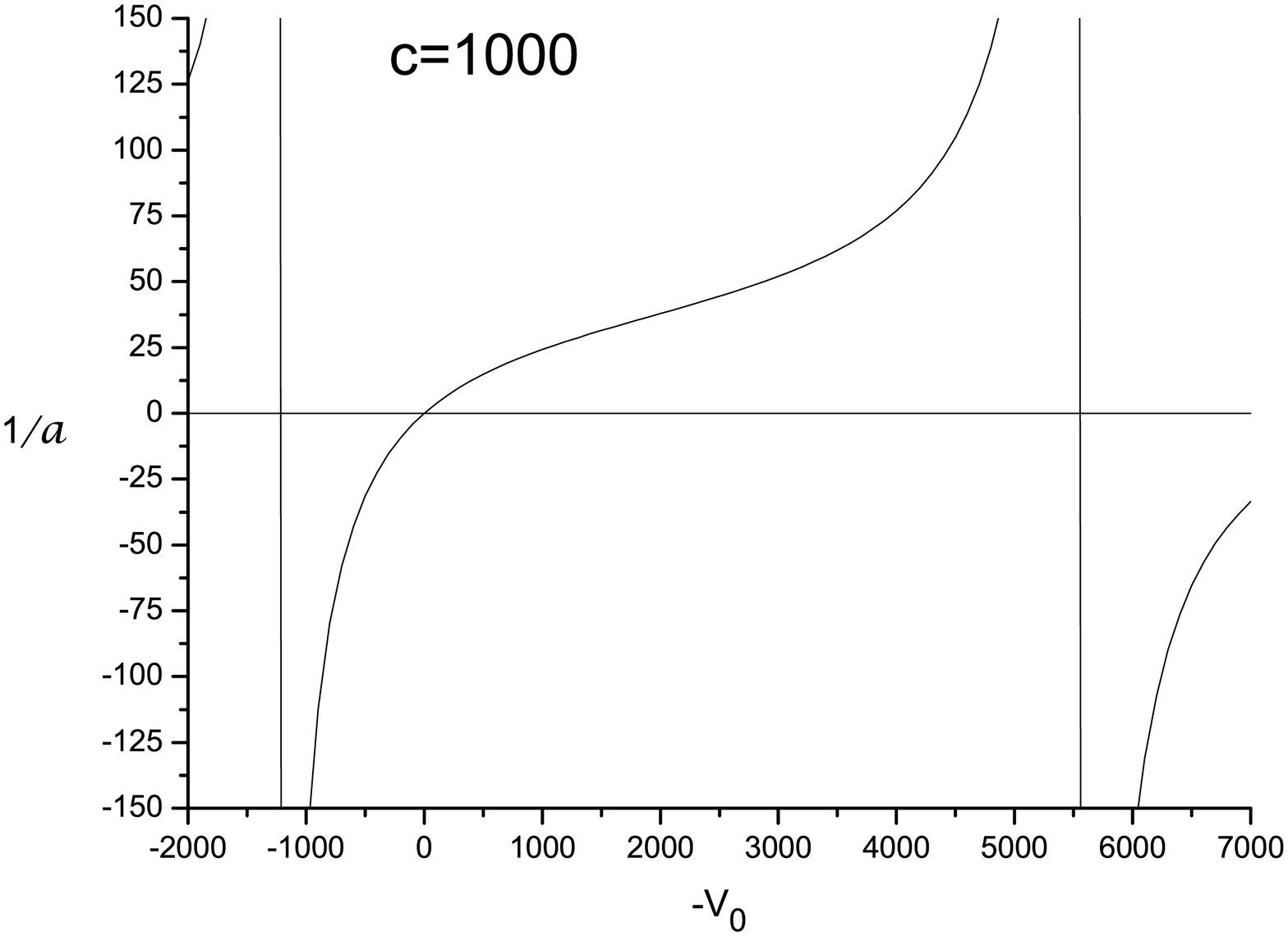} \\[-2.5mm]

\end{tabular}
\end{center}
Рис. 1: Зависимость обратной длины рассеяния $ a^{-1} $ от глубины $V_0$ потенциальной ямы $V(x)=V_0\exp{\left\{ -2cx^2\right\}}$ и параметра $c$
\\[+4.5mm]

Для численного решения задачи на собственные значения (1) использовался метод обратной итерации. Для аппроксимации производных, входящих в УШ (1), были использованы конечно-разностные аппроксимации второго порядка точности, обеспечивающие погрешность порядка $h^2$, где $h$ - есть шаг разностной сетки. Результаты расчета представлены на Рис. 2-6, где также представлены рассчитанные волновые функции. Один из основных результатов статьи представлен на Рис.2.

\begin{center}
\begin{tabular}{c}

\includegraphics*[width=160mm]{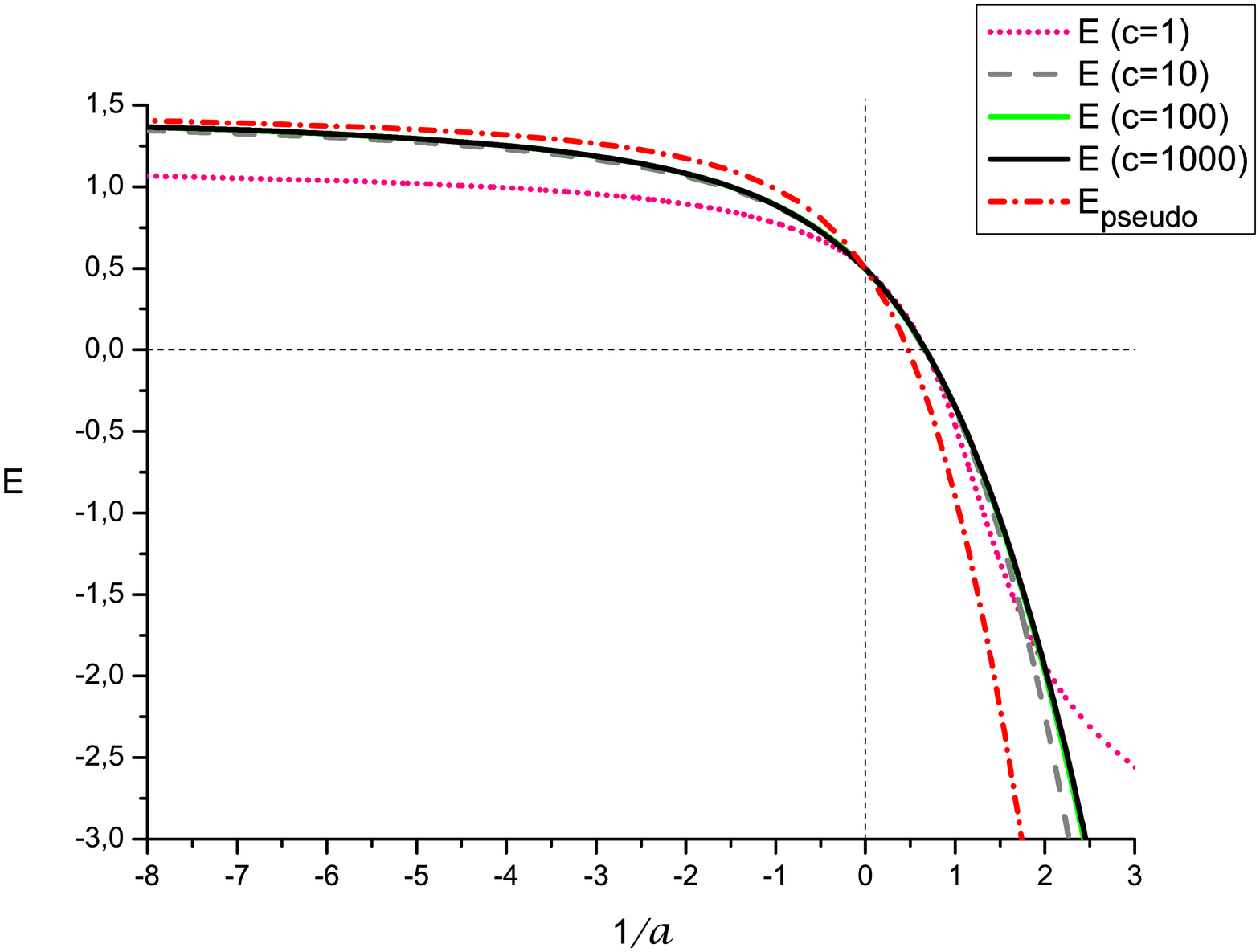}

\end{tabular}
\end{center}
Рис. 2: Энергия основного состояния при различных $ c $; $E_{\textrm{pseudo}}$ - расчет сделанный по формуле из работы \cite{Busch}
\\[+4.5mm]

Из графика видно, что при одновременном сужении и углублении ($c,V_0\rightarrow \infty$) потенциала взаимодействия двух атомов, т.е. при переходе к короткодействующему взаимодействию, в представленном на графике диапазоне $1/a$, мы получаем сходящийся результат\footnote[1]{в данном диапазоне $1/a$ различие между $c=100$ и $c=1000$ едва заметно.}. При сравнении этого результата с результатом работы \cite{Busch} мы обнаруживаем не совпадение кривых, которое растет в области положительных значений $1/a$. В пределе $c\rightarrow\infty$ гауссов потенциал качественно стремится к потенциалу (псевдопотенциалу) нулевого радиуса, использованному в работе \cite{Busch}. Из общих соображений, однако,  ясно, что атомы обладают структурой, поэтому потенциал должен иметь некий радиус действия.

\section{Метод осцилляторного представления (ОП)}

Уравнение Шредингера (1) также решалось с помощью метода ОП. Этот подход основан на методах и идеях скалярной квантовой теории поля \cite{Efimov}. Рассмотрим гамильтониан скалярного поля (в системе $\hbar=c=1$):
\begin{eqnarray}
H=H_0+H_I=\frac{1}{2}\int dx\left[ \pi^2(x)+(\nabla \phi(x))^2+m^2\phi^2(x) \right]+g\int dx\phi^4(x)
\end{eqnarray}
Квадратичный "свободный" гамильтониан $H_0$ описывает основное состояние, если константа связи $g$ мала. Однако в формировании основного состояния, возникающему благодаря взаимодействию $H_I$ при любой величине константы связи $g$, главную роль играют так называемые "диаграммы-головастики" (или диаграммы "кактусного" типа) и в квантовой теории поля они составляет наибольшую расходимость. Чтобы можно было описать основное состояние системы полевые операторы представляют через операторы рождения и уничтожения и упорядочивают их в нормальной форме. Требование чтобы гамильтониан взаимодействия, представленный в нормальной форме, содержал только полевые операторы степени три и выше
означает перенормировку массы скалярной частицы $m$ и энергии вакуума, т.е. учету диаграмм Фейнмана "кактусного" типа. Таким образом, после того как гамильтониан системы представлен в нормальной форме, его квадратичная часть, соответствующая свободному гамильтониану, описывает основное состояние, а члены третьей степени и выше описывают необходимые поправки.

Процедура решения следующая. Пусть в одномерном пространстве $\mathbb{R}^1$ задан гамильтониан системы вида:
\begin{eqnarray}
H=\frac{p^2}{2}+W(q^2),
\end{eqnarray}
где $W(q^2)$ - есть потенциальная энергия. Выделяем чисто осцилляторную часть:
\begin{eqnarray}
H=\frac{1}{2}\left( p^2 +\Omega^2 q^2 \right)
+\left[ W(q^2) - \frac{\Omega^2}{2}q^2 \right],
\end{eqnarray}
где $\Omega$ является пока произвольным положительным параметром. Канонические переменные $p$,$q$ представим через операторы рождения и уничтожения:
\begin{eqnarray}
\nonumber
q&=&\frac{a+a^+}{\sqrt{2\Omega}} \\ \nonumber \\
p&=&\sqrt{\frac{\Omega}{2}}\frac{a-a^+}{i}
\end{eqnarray}
Далее подставляем выражения для канонических переменных $q$ и $p$ в (6) и производим упорядочивание по операторам рождения и уничтожения. Тогда потенциал, представленный в нормальной форме, принимает вид:
\begin{eqnarray}
W(q^2)=\int \left( \frac{dk}{2\pi} \right) \tilde{W}(k^2)\exp{ \left\{-\frac{k^2}{4\Omega} \right\} }
:\exp{ \left\{ ikq \right\} }:,
\end{eqnarray}
где $:*:$  - есть символ нормального упорядочения и
\begin{eqnarray}
\tilde{W}(k^2)=\int d\rho W(\rho^2) e^{ik\rho}
\end{eqnarray}
Таким образом, имеем:
\begin{eqnarray}
\nonumber
&\dfrac{1}{2}\left( p^2+\Omega^2q^2 \right)=\Omega (a^+ a)+\dfrac{1}{2}\Omega& \\ \nonumber \\
&W(q^2)-\dfrac{\Omega^2}{2}q^2=
\int \left( \dfrac{dk}{2\pi} \right) \tilde{W}(k^2)\exp{ \left\{-\dfrac{k^2}{4\Omega} \right\} }
:\exp{ \left\{ ikq \right\} }:
-\dfrac{\Omega^2}{2}\left( :q^2:+\dfrac{1}{2\Omega} \right)&
\end{eqnarray}

Потребуем, чтобы гамильтониан взаимодействия, представленный в нормальной упорядоченной форме, не содержал каких-либо слагаемых, квадратичных по каноническим переменным, поскольку предполагается, что квадратичные члены определяют осцилляторный характер взаимодействия и полностью включены в свободный гамильтониан $\Omega (a^+ a)$. Это требование будем называть \textit{условием осцилляторного представления}. Отсюда возникает возможность получить уравнение для частоты осциллятора $\Omega$:
\begin{eqnarray}
\Omega^2+
\int \left( \dfrac{dk}{2\pi} \right) \tilde{W}(k^2)\exp{ \left\{-\dfrac{k^2}{4\Omega} \right\} }k^2=0
\end{eqnarray}
Используя эти соотношения, перепишем гамильтониан (6) в виде:
\begin{eqnarray}
H=H_0+H_I+\varepsilon_0,
\end{eqnarray}
где
\begin{eqnarray}
\nonumber
&H_0=\Omega (a^+a),& \\ \nonumber \\ \nonumber
&H_I=\int \left( \dfrac{dk}{2\pi} \right) \tilde{W}(k^2)\exp{ \left\{-\dfrac{k^2}{4\Omega} \right\} }
:e^{ikq}_2:,& \\ \nonumber \\ \nonumber
&\varepsilon_0=\dfrac{\Omega}{4}+\int \left( \dfrac{dk}{2\pi} \right) \tilde{W}(k^2)\exp{ \left\{-\dfrac{k^2}{4\Omega} \right\} },& \\ \nonumber \\ \nonumber
&e^z_2=e^z-1-z-\frac{z^2}{2},&
\end{eqnarray}
а $\varepsilon_0$ - есть энергия основного состояния или вакуума гамильтониана (6). Из (11) видно, что уравнение для $\Omega$ полученное из условия осцилляторного представления, совпадает с уравнением, определяющим минимум энергии $\varepsilon_0$ (12) по $\Omega$, т.е.
\begin{eqnarray}
\nonumber
\varepsilon_0&=&\min\limits_{\Omega}\left\{ \dfrac{\Omega}{4}+\int \left( \dfrac{dk}{2\pi} \right) \tilde{W}(k^2)\exp{ \left\{-\dfrac{k^2}{4\Omega} \right\} } \right\}= \\ \nonumber  \\ \nonumber
&=&\min\limits_{\Omega}\left\{  \dfrac{\Omega}{4}+\int \left( \dfrac{d\rho}{\sqrt{\pi}} \right) W\left( \dfrac{\rho^2}{\Omega} \right) \exp{ \left\{- \rho^2 \right\} }  \right\}= \\ \nonumber  \\ \nonumber
&=&\min\limits_{\Omega}\left\{  \dfrac{\Omega}{4}+\int_0^{\infty} \dfrac{du \cdot e^{-u}}{\sqrt{\pi u}} W\left( \dfrac{u}{\Omega} \right)            \right\}= \\ \nonumber  \\
&=& \int_0^{\infty} \dfrac{du \cdot e^{-u}}{\sqrt{\pi u}} \dfrac{d}{du} \left[ u W\left( \dfrac{u}{\Omega} \right) \right],
\end{eqnarray}
где $\Omega$ вычисляется из (11), которое можно переписать в виде
\begin{eqnarray}
\Omega =\dfrac{2}{\sqrt{\pi}} \int_0^{\infty} du \cdot \sqrt{u} e^{-u} \dfrac{d}{du} W\left( \dfrac{u}{\Omega} \right).
\end{eqnarray}
Таким образом, частота осциллятора $\Omega$ есть функция исходных параметров потенциала $W \left( q^2 \right)$.

Проделав описанную выше процедуру формализма ОП для конкретного гамильтониана (1) мы придем к следующей оптимизационной задаче\footnote[1]{формула (15) выведена Г.В. Ефимовым}(в системе $m=\hbar=\omega=1$):
\begin{eqnarray}
\nonumber
&\varepsilon_0=\min\limits_{\Omega}\left\{ \dfrac{\Omega}{4}+\dfrac{1}{4\Omega}
+\dfrac{V_0}{ \sqrt{1+\dfrac{2c}{\Omega}} } \right\}& \\ \nonumber \\
&\Omega^2=1-\dfrac{4 c V_0}{ \left(  1+\dfrac{2c}{\Omega}    \right)^{3/2} }&
\end{eqnarray}

Выделение в исходной задаче гамильтониана взаимодействия $H_I$ позволяет в рамках теории возмущений (ТВ) вычислять высшие поправки к энергии уровня. Первый порядок ТВ в ОП равен нулю. Второй порядок вычисляется по формуле \cite{Efimov}:

\begin{eqnarray}
\varepsilon_2 = -\frac{1}{2\Omega} \sum_{n=2}^{\infty}B\left( \dfrac{1}{2}, \dfrac{1}{2}+n\right)  A_n^2,
\end{eqnarray}
где $\Gamma(x), B(x)$ - есть гамма и бета функции соответственно,
\begin{eqnarray}
A_n=\dfrac{V_0}{n! \sqrt{\pi}} \left( \dfrac{2c}{\Omega}\right)^n\dfrac{\Gamma \left( n+\dfrac{1}{2} \right) }
{\left(1+\dfrac{2c}{\Omega} \right)^{n+1/2}}
\end{eqnarray}
Таким образом, энергия основного состояния исходного гамильтониана (5), или (1), вычисляется как
\begin{eqnarray}
E_0=\varepsilon_0+\varepsilon_2+\ldots
\end{eqnarray}

Результаты расчетов методом ОП при различных параметрах потенциала взаимодействия представлены на Рис. 3-6. На графиках также представлены кривые соответствующие второму порядку теории возмущений метода ОП.
\begin{center}
\begin{tabular}{ccc}
\includegraphics*[width=90mm]{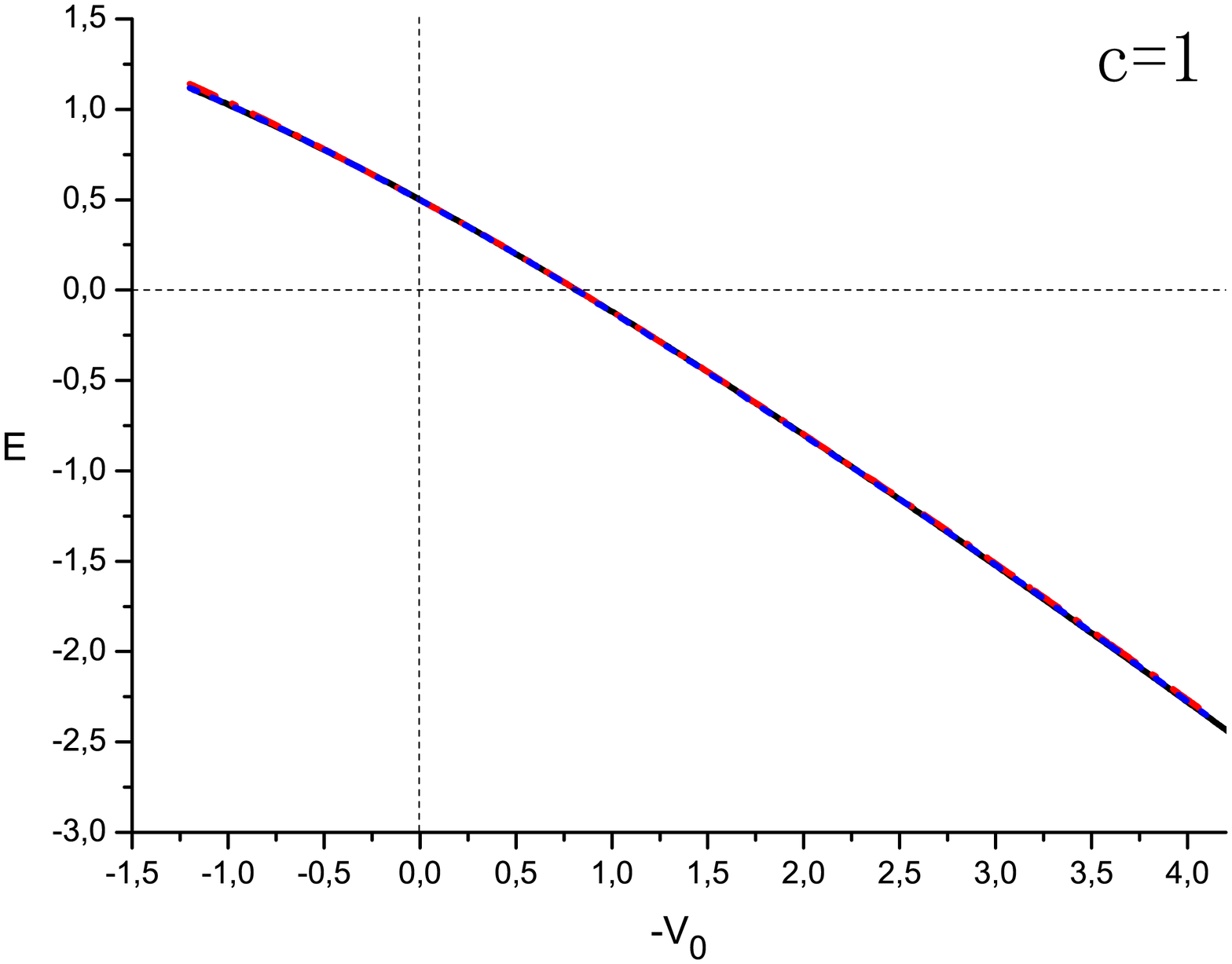} & \hspace{-16mm}
\includegraphics*[width=90mm]{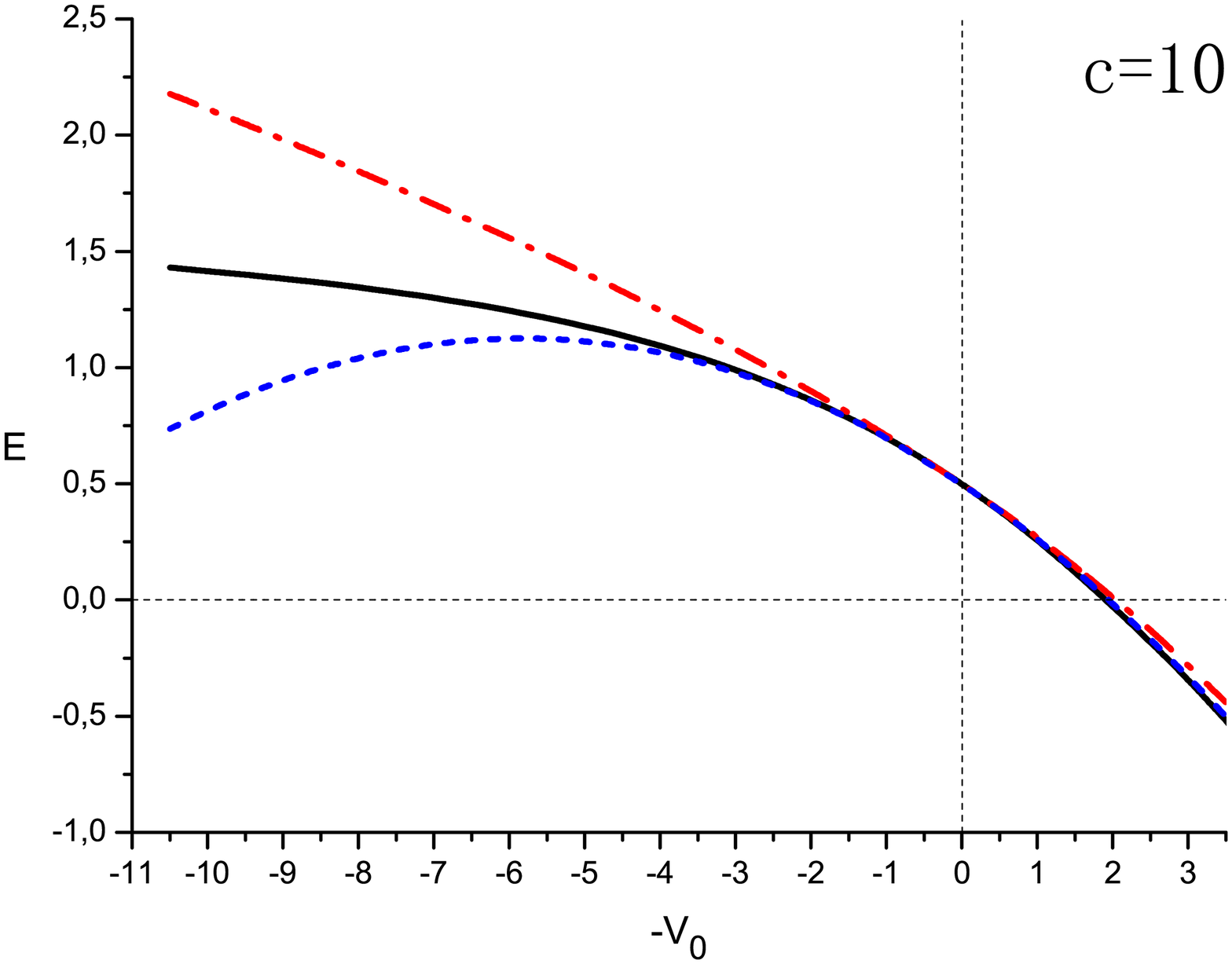} \\[+1mm]

\includegraphics*[width=90mm]{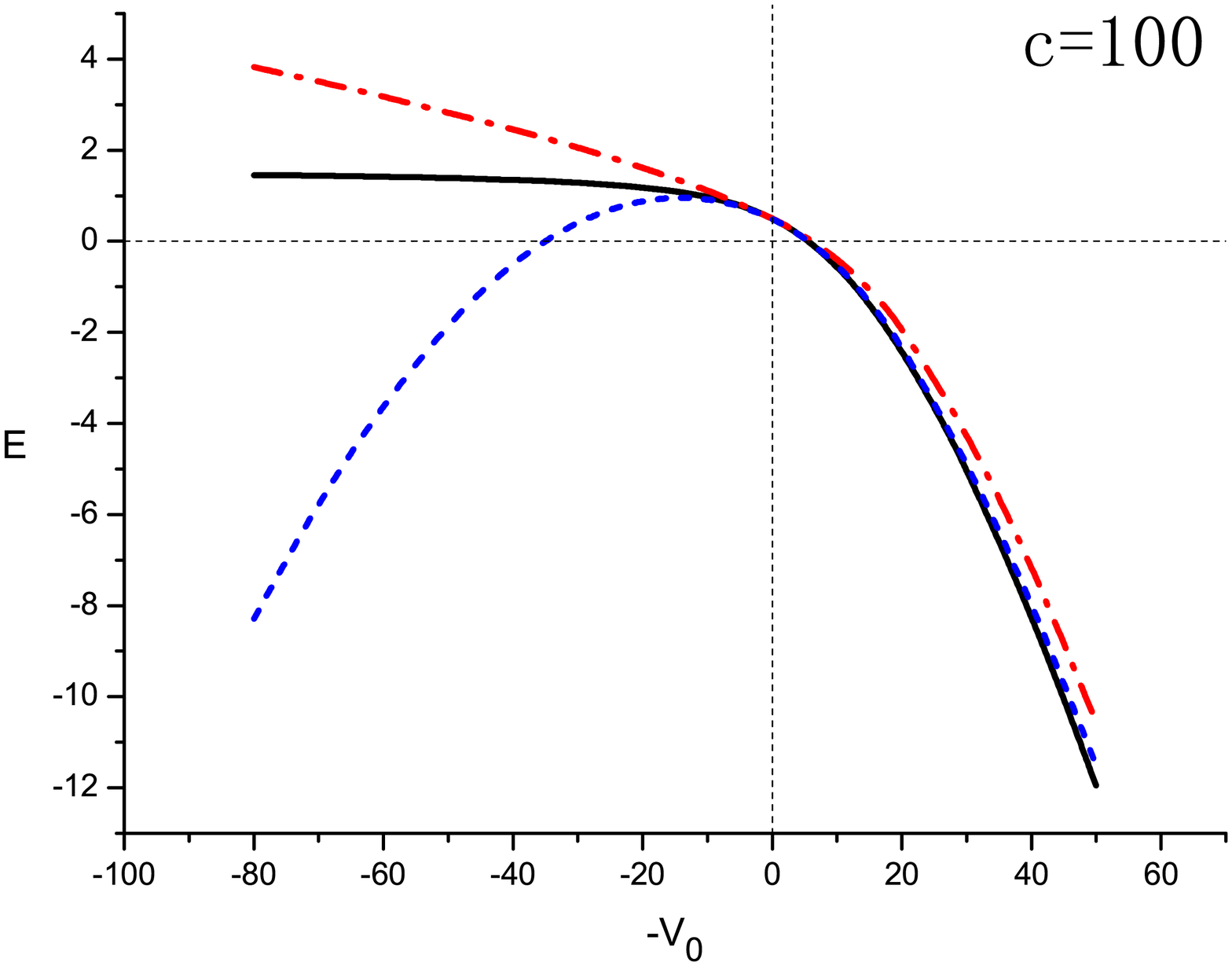} & \hspace{-16mm}
\includegraphics*[width=90mm]{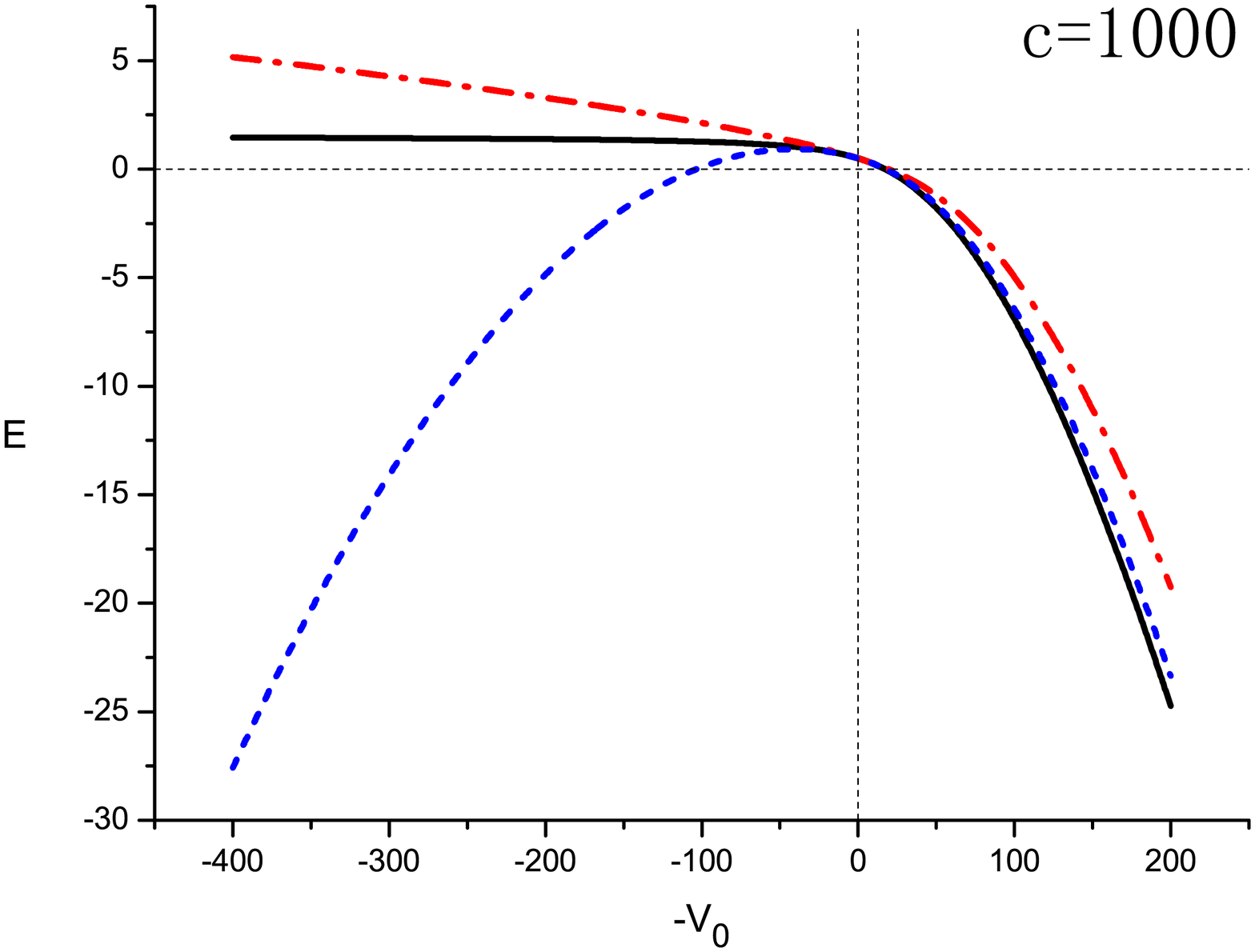} \\
\end{tabular}
\end{center}
Рис. 3: Границы применимости метода ОП для уравнения (1);  Сплошные линии - численный расчет энергии. Штрихпунктирные линии - энергия в нулевом приближении ОП. Пунктирные линии - энергия во втором приближении ОП.
\\[+4.5mm]

\begin{center}
\begin{tabular}{c}

\includegraphics*[width=160mm]{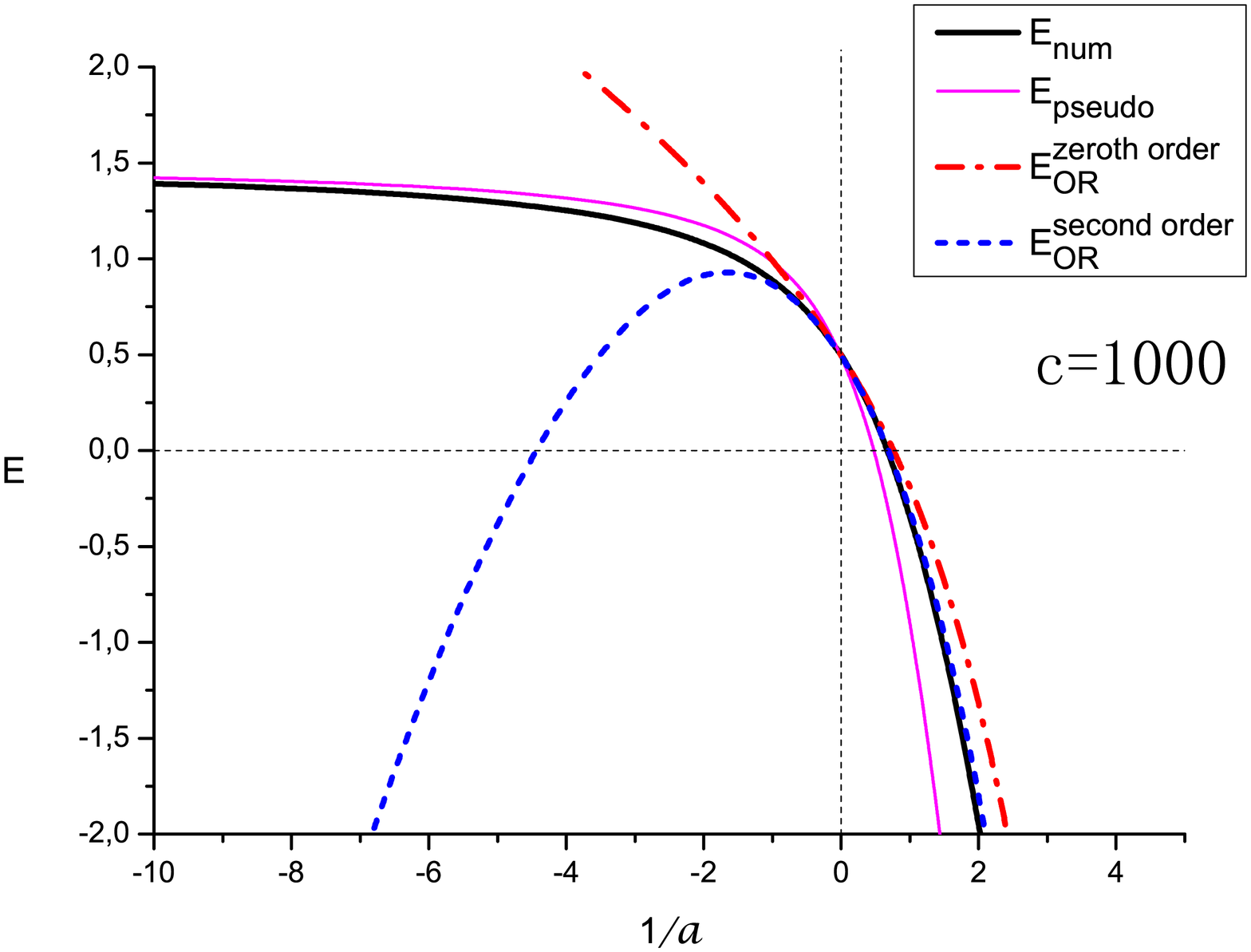}

\end{tabular}
\end{center}
Рис. 4: Границы применимости метода ОП для уравнения (1); $E_{\textrm{num}}$ - численный расчет энергии, $E_{\textrm{pseudo}}$ - энергия по формуле из работы \cite{Busch}, $E_{\textrm{OR}}^{\textrm{zeroth order}}$ - энергия в нулевом приближении ОП, $E_{\textrm{OR}}^{\textrm{second order}}$ - энергия во втором приближении ОП.
\\[+4.5mm]

Из графиков видно, что при отрицательных значениях $V_0<0$ метод ОП начинает отклоняться от численных расчетов при $c,V_0 \rightarrow \infty$. Второй порядок ТВ в ОП отклоняется при этих же параметрах потенциала меньше и в области отрицательных $V_0$ показывает очень хорошую точность. Однако, для положительных $V_0>0$ как нулевой, так и второй порядок ТВ метода ОП при $c,V_0 \rightarrow \infty$ отклоняются от численного расчета значительно.

Причина отклонения метода ОП при различных параметрах потенциала объясняется тем, что в ОП волновая функция, в нулевом приближении, описывается гауссианом \cite{Efimov}:
 \begin{eqnarray}
 \Psi_{OR}(x)=\left( \frac{\Omega}{\pi} \right)^{1/4}\exp{ \left\{ -\frac{\Omega}{2}x^2 \right\} }
 \end{eqnarray}
Тем самым отклонение ОП обусловлено отклонением истинной ВФ от гауссова поведения осцилляторной волновой функции, что можно пронаблюдать на Рис. 5,6.

\begin{center}
\begin{tabular}{cc}

\includegraphics*[width=80mm]{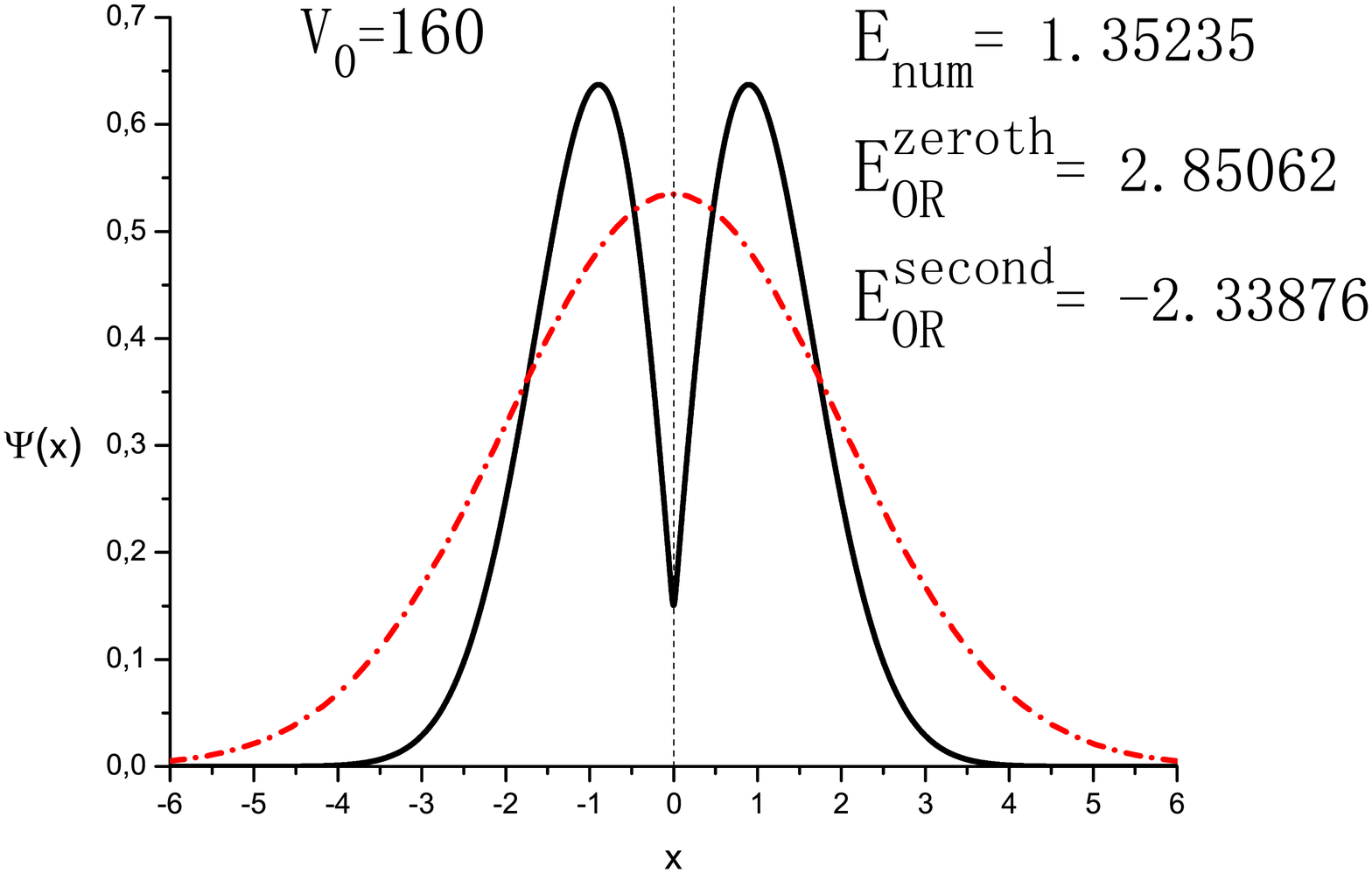} &
\includegraphics*[width=80mm]{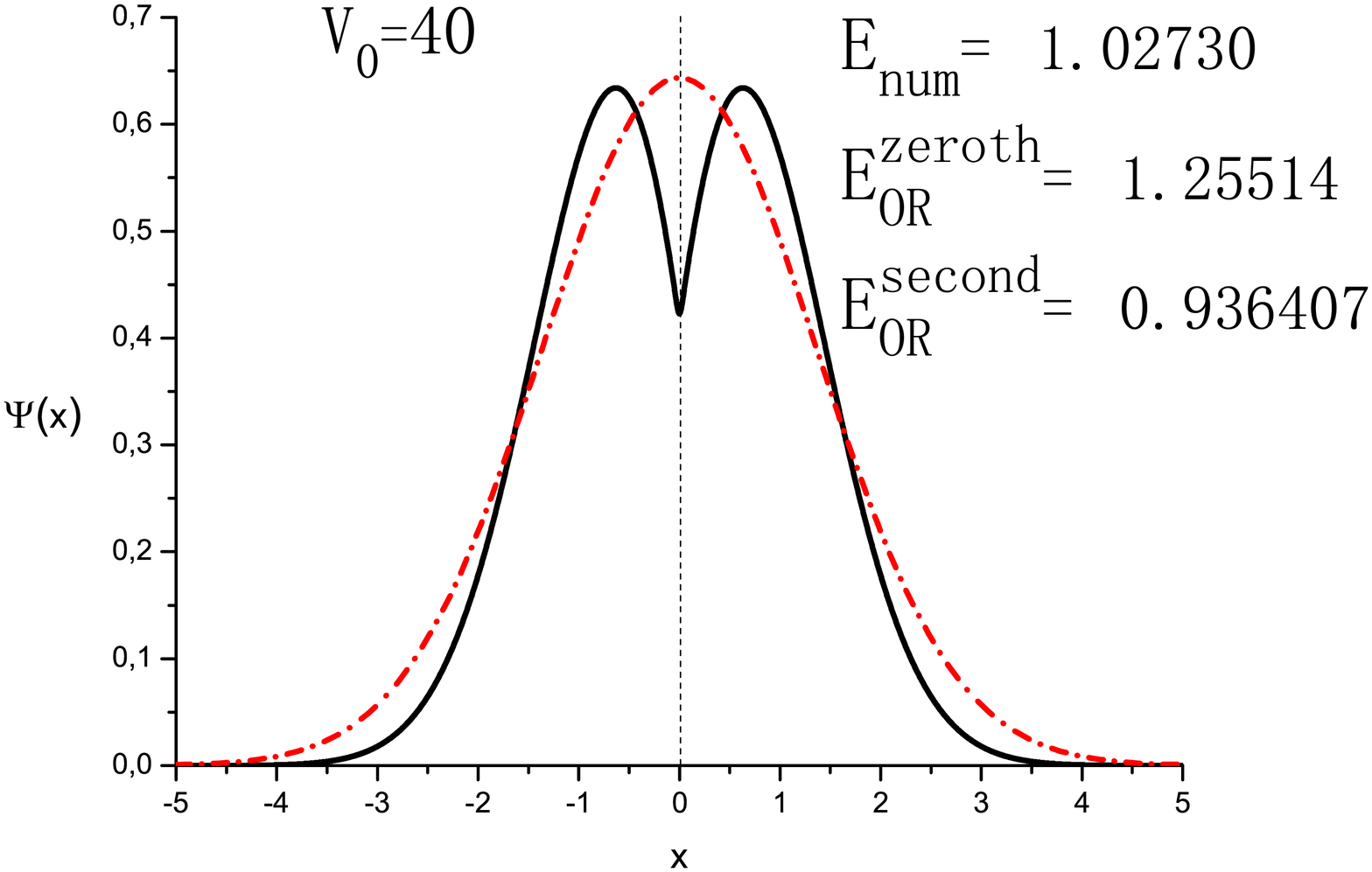} \\[-2.5mm]

\includegraphics*[width=80mm]{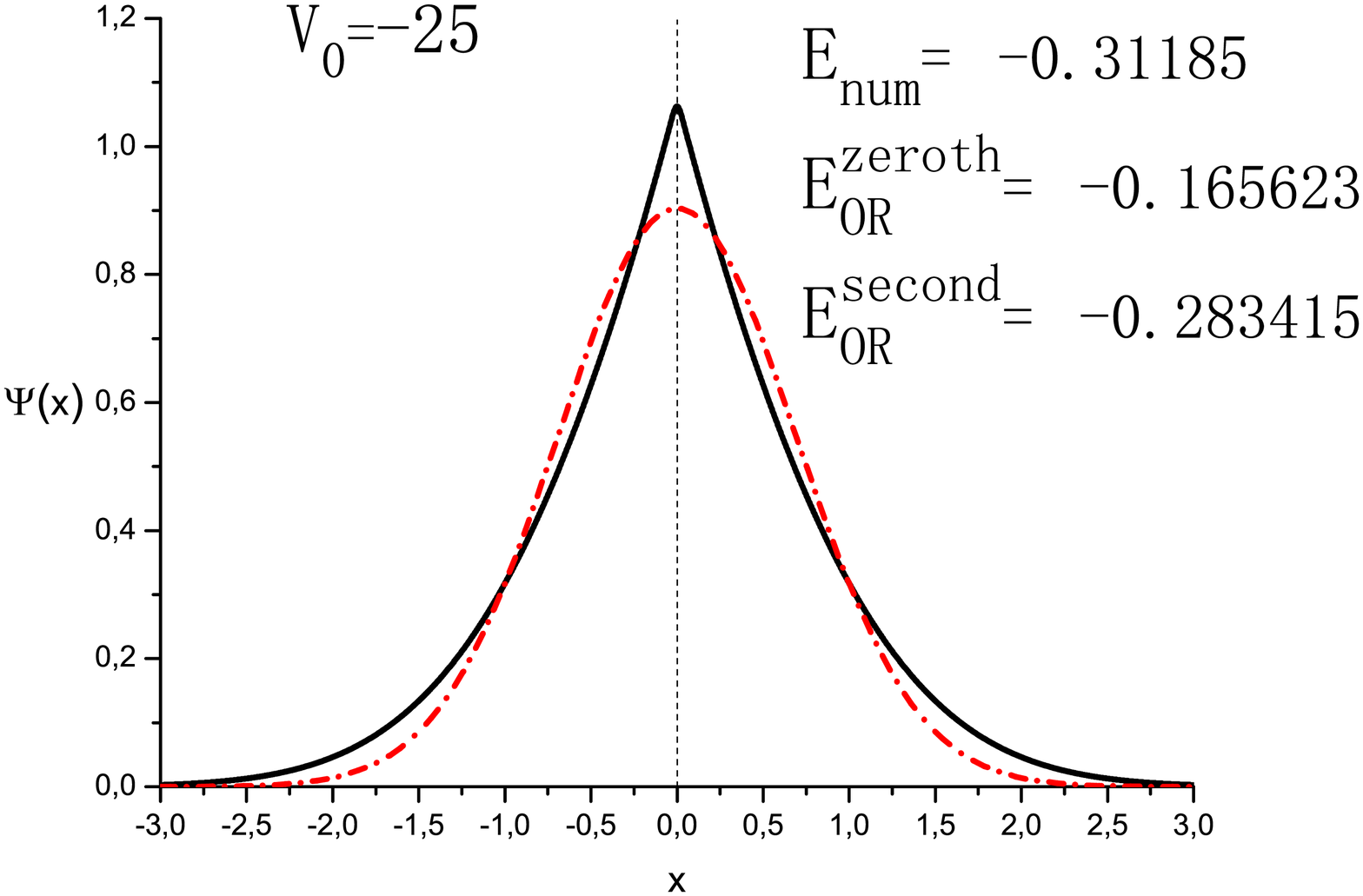} &
\includegraphics*[width=80mm]{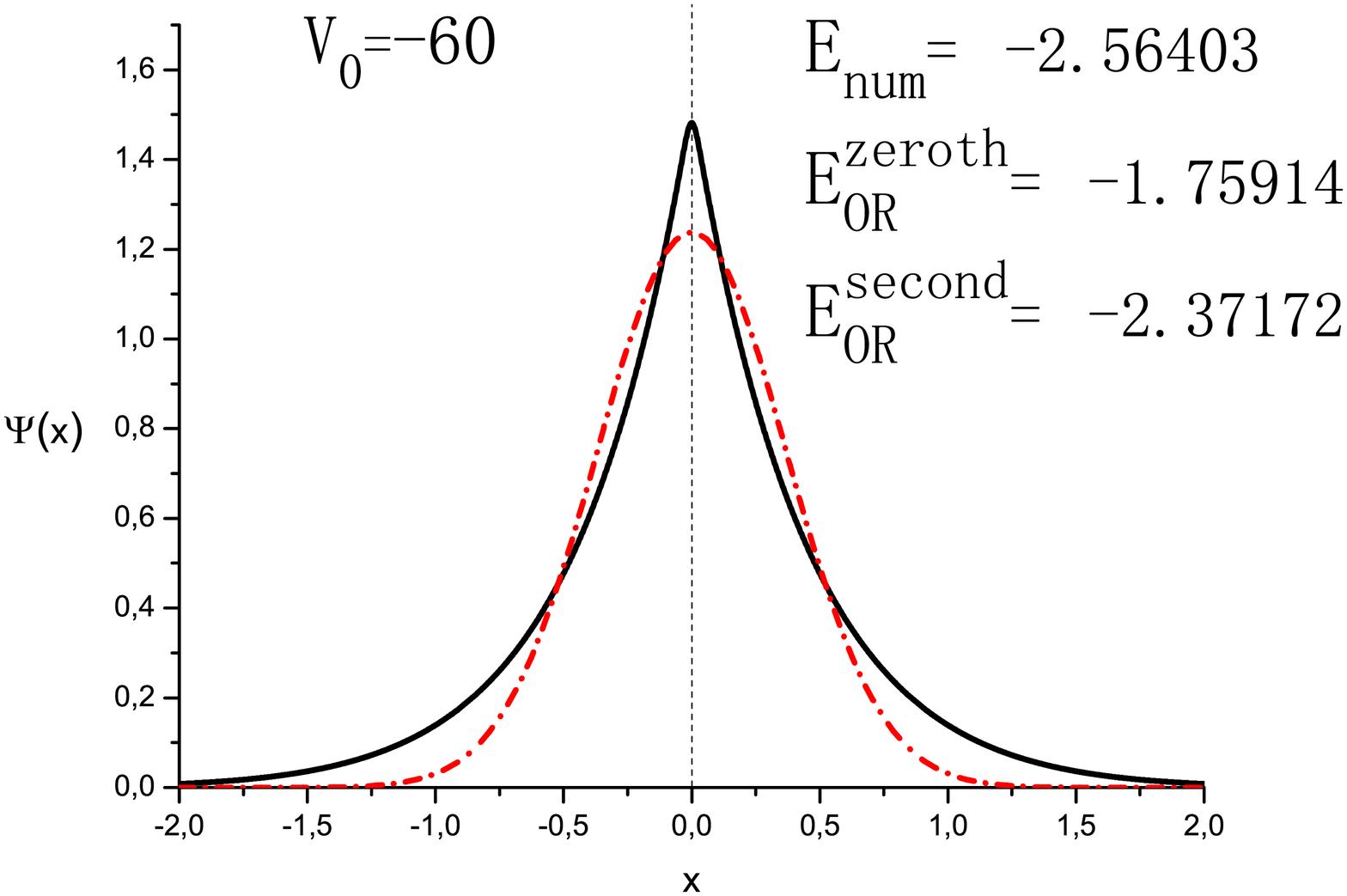} \\[-2.5mm]

\end{tabular}
\end{center}
Рис. 5: Сравнение ВФ численного расчета и ВФ метода ОП при различных $V_0$ для $c=1000$. Сплошные линии - ВФ численного расчета. Штрихпунктирные линии - ВФ нулевого порядка ОП. $E_{\textrm{num}}$ - численный расчет энергии, $E_{\textrm{OR}}^{\textrm{zeroth}}$ - энергия в нулевом приближении ОП, $E_{\textrm{OR}}^{\textrm{second}}$ - энергия во втором приближении ОП.
\\[+4.5mm]

\begin{center}
\begin{tabular}{cc}
\includegraphics*[width=80mm]{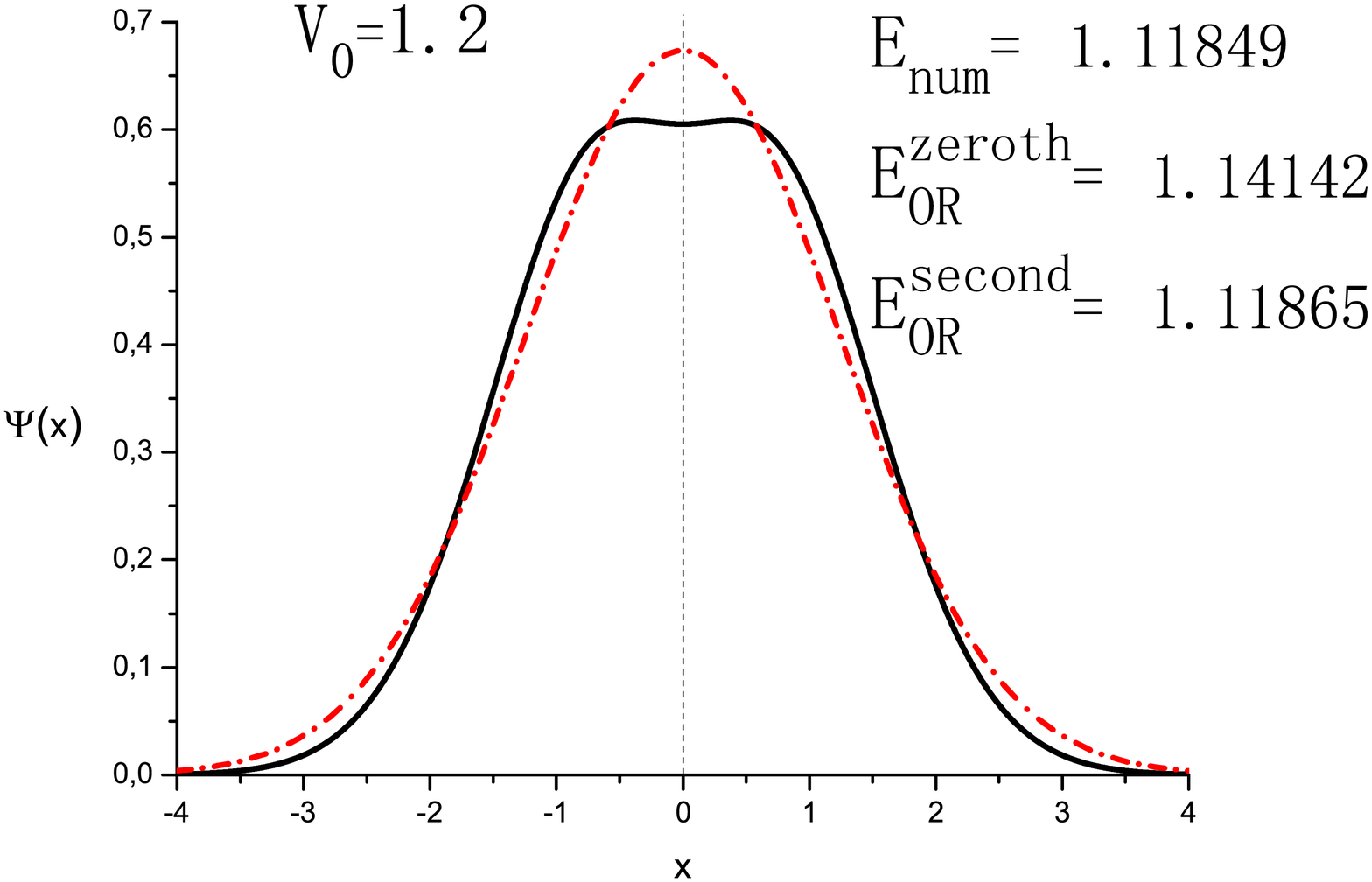} &
\includegraphics*[width=80mm]{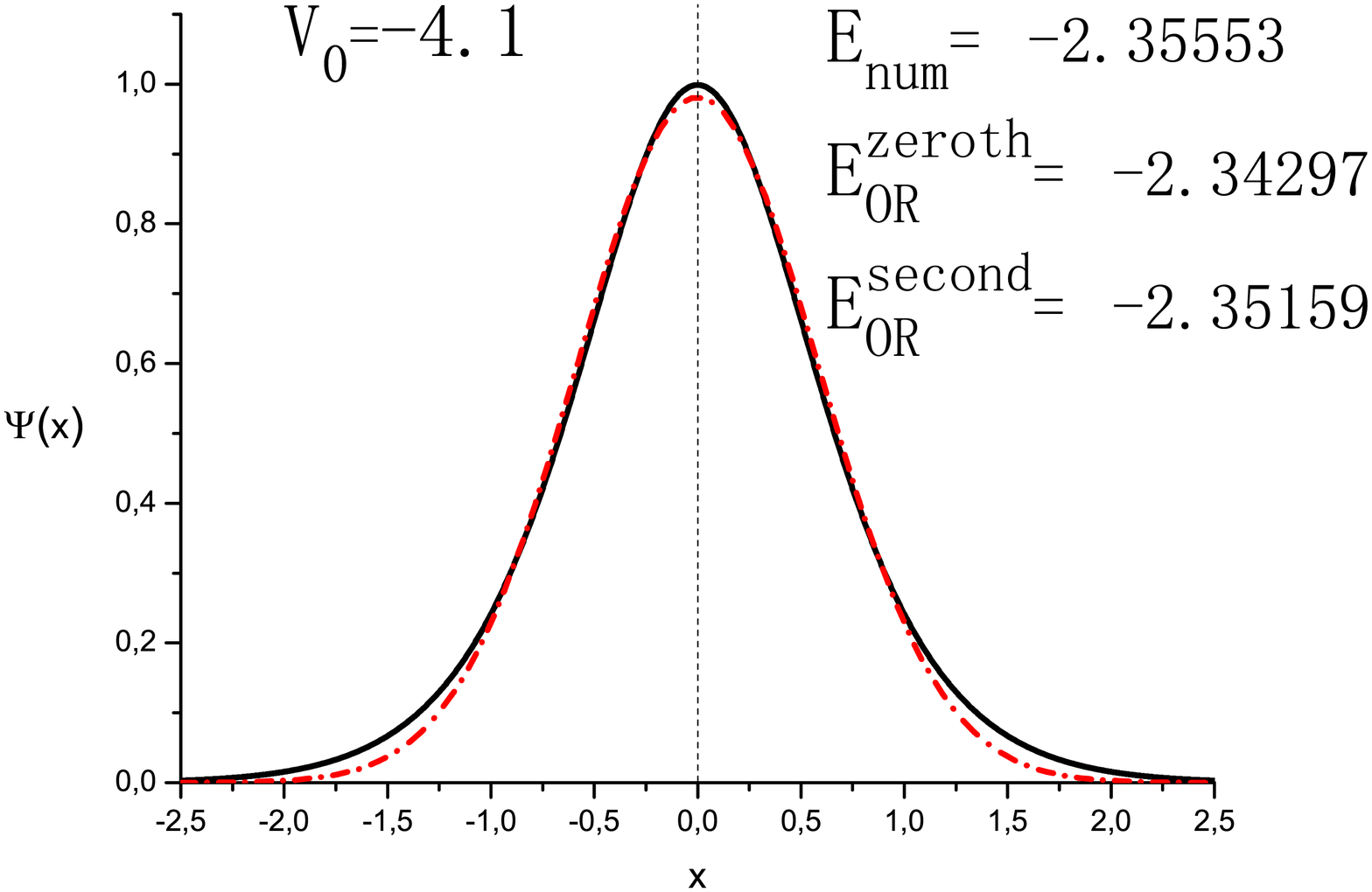} \\[-2.5mm]
\end{tabular}
\end{center}
Рис. 6: Сравнение ВФ численного расчета и ВФ метода ОП при различных $V_0$ для $c=1$. Сплошные линии - ВФ численного расчета. Штрихпунктирные линии - ВФ нулевого порядка ОП. $E_{\textrm{num}}$ - численный расчет энергии, $E_{\textrm{OR}}^{\textrm{zeroth}}$ - энергия в нулевом приближении ОП, $E_{\textrm{OR}}^{\textrm{second}}$ - энергия во втором приближении ОП.
\\[+4.5mm]

Видно, что отклонение ОП от численного расчета, возникающее в области положительных $V_0$, при $c,V_0 \rightarrow \infty$, соответствует отклонению волновых функций этих двух подходов. Таким образом можно утверждать, что для данного гамильтониана (1) применимость метода ОП является функцией параметров потенциала взаимодействия.
\section{Заключение}
В данной работе была рассчитана энергия основного состояния УШ (1), описывающего двухбозонную связанную систему в одномерной  гармонической ловушке, как функция обратной длины рассеяния. Полученная зависимость $E(a^{-1})$ несколько отличается от расчета работы \cite{Busch}, выполненного в псевдопотенциальном подходе. Наш случай отличается тем, что мы использовали реалистичный потенциал гаусса, выбор которого является более естественным, так как он учитывает наличие у атомов некой структуры. Также можно заключить то, что переход от
потенциала конечного радиуса к потенциалу дельта-функции является существенным приближением. Похожее сравнение расчетов выполнены с трехмерными реалистическим потенциалом и псевдопотенциалом нулевого радиуса в \cite{Olshanii} для двумерной гармонической ловушки.

Определена область применимости метода осцилляторного представления для задачи (1) как функция параметров потенциала взаимодействия. Отклонение метода ОП от численного расчета обусловлено отклонением истинной ВФ от гауссова поведения осцилляторной ВФ. Определение области применимости ОП необходимо при дальнейшем применении метода в случае расчета спектра двухатомной системы в ангармонической ловушке.

Основная сложность при расчете спектра двухатомной системы в ангармонической ловушке связана с тем, что в этом случае разделение переменных относительной координаты и координаты центра масс уже не представляется возможным. Возможность расчета альтернативными методами будет способствовать надежности результатов.

Авторы выражают благодарность В.С. Мележику и Г.В. Ефимову: В.С. Мележику - за постановку задачи, за связанные с ней дискуссии и за проверку правильности ее выполнения; Г.В. Ефимову - за дискуссии связанные с работой \cite{Busch} и за применение метода ОП.

\end{document}